\documentclass[preprint,showpacs,groupedaddress,superscriptaddress]{revtex4}

\def\G{\mathord{\buildrel{\lower3pt\hbox{$\scriptscriptstyle\leftrightarrow$}}
\over G}}

\usepackage{epsf,epsfig}
\usepackage[psamsfonts]{amssymb}
\usepackage{amsmath}
\usepackage{bm}
\begin{document}
\title{Finite temperature Casimir effect in the presence of nonlinear dielectrics}
\author{Fardin Kheirandish}
\email[]{fardin_kh@phys.ui.ac.ir} \affiliation{Department of
Physics, Faculty of Science, University of Isfahan, Isfahan,
Iran.}
\author{Ehsan Amooghorban}
\email[]{amooghorban@sci.ui.ac.ir} \affiliation{Department of
Physics, Faculty of Science, University of Isfahan, Isfahan,
Iran.}
\author{Morteza Soltani}
\email[]{msoltani@phys.ui.ac.ir} \affiliation{Department of
Physics, Faculty of Science, University of Isfahan, Isfahan,
Iran.}

\begin{abstract}
\noindent Starting from a Lagrangian, electromagnetic field in the
presence of a nonlinear dielectric medium is quantized using
path-integral techniques and correlation functions of different
fields are calculated. The susceptibilities of the nonlinear
medium are obtained and their relation to coupling functions are
determined. Finally, the Casimir energy and force in the presence
of a nonlinear medium at finite temperature is calculated.
\end{abstract}
\maketitle
\section{Introduction}
\noindent One of the most direct manifestations of the zero-point
vacuum oscillations is the Casimir effect. This effect in its
simplest form, is the attraction force between two neutral,
infinitely large, parallel and ideal conductors in vacuum. The
effect is completely quantum mechanical and is a result of
electromagnetic field quantization in the presence of some
boundary conditions. The presence or absence of boundary
conditions cause a finite change of vacuum-energy which its
variation respect to the distance between the conductors gives the
Casimir force [1]. The general theory of forces between parallel
dielectrics was worked out by Lifshitz et al \cite{2}. Some years
later, the theory of the Casimir effect was rederived by Schwinger
\cite{3} and later on this theory extended to systems consisting
of some dielectric layers \cite{4}-\cite{5} and precision
experiments on measuring the Casimir force performed
\cite{6}-\cite{9}.
\par

Generally, calculations of Casimir forces are based on two
different approaches \cite{10}. In the first approach, the total
energy of discrete quantum modes of the electromagnetic field is
calculated and usually a regularization procedure is required,
although there is still no consensus on the regularization
procedures, which often lead to inconsistent results. On the other
hand, this formalism despite of its simplicity, is restricted to
systems where energy eigenvalues are known, but geometries in
which energy eigenvalues are known exactly are few.\par

The second approach is based on a Green's function method where
using the fluctuation$–$dissipation theorem, the electromagnetic
field energy density is linked directly to the photonic Green's
function. \par

In recent years considerable attention has paid to the Casimir
effect because of its wide applications in different areas of
physics such as quantum field theory, gravitation and cosmology,
atomic physics, condensed matter, nanotechnology, and mathematical
physics \cite{11}-\cite{14}. On the other hand, during the last
two decades, the emergence of periodically structured optical
materials—commonly called photonic crystals- has lead to
substantial progress in the science and technology of optics and
photonics \cite{15}. Photonic crystals \cite{16}-\cite{17} provide
new opportunities for enhancing and controlling nonlinear optical
processes. An important step in this direction is to provide a
quantum theory of light that takes into account nonlinear
processes and at the same time include the absorption in the
nonlinear material. Several attempts have been made to provide a
quantized theory of macroscopic nonlinear electrodynamics
\cite{18}-\cite{23} and most of them are focused on strictly
lossless materials. Recently, a consistent approach that includes
absorption and dispersion has been formulated within the frame
work of quantum brownian motion \cite{24}. This motivated us to
investigate the quantization of electromagnetic field in the
presence of a nonlinear dielectric medium and also consider the
nonlinear effects of the medium on Casimir effect which can have
applications in both fundamental science and engineering. \par

In the 1990s, Golestanian and Kardar developed a path-integral
approach to investigate the dynamic Casimir effect in a system
which was consisted of two corrugated conducting plates surrounded
by quantum vacuum \cite{25}-\cite{26}. Emig and his colleagues
also used the path-integral formalism to obtain normal and lateral
Casimir force between two sinusoidal corrugated perfect conductor
surfaces \cite{27}-\cite{28}. Recently, this formalism is merged
with a canonical approach to calculate the Casimir force between
two perfectly conducting plates immersed in a magnetodielectric
medium \cite{29}. In the present work the electromagnetic field in
the presence of a nonlinear magnetodielectric medium is quantized
in the framework of path-integrals and as an example the mutual
Casimir force between conducting parallel plates, immersed in a
nonlinear dielectric medium, is calculated.\par

The layout of the present work is as follows: In Sec.II, a
Lagrangian for electromagnetic field in the presence of a
nonlinear medium is proposed and quantization is achieved via
path-integrals. In Secs.III and IV, the linear and nonlinear
green's functions of the electromagnetic field and medium are
obtained and the nth order susceptibility of the nonlinear medium
which satisfies the kramers kronig relations is determined. In
Sec.V, Casimir force in the presence of a linear and nonlinear
medium for zero and finite temperature is calculated. Finally, we
discuss the main results and conclude in Sec.VI.

\section{Field quantization}
\noindent Quantum electrodynamics in a linear magnetodielectric
medium can be accomplished by modeling the medium with two
independent reservoirs interacting with the electromagnetic field.
Each reservoir contains a continuum of three dimensional harmonic
oscillators describing the electric and magnetic properties of the
medium \cite{30}-\cite{32}. In this section we follow the idea
introduced in \cite{24} to quantize the electromagnetic field in
the presence of a nonlinear medium and for simplicity we restrict
our attention to a nonmagnetic medium but the generalization to a
magnetodielectric medium is straightforward. For this purpose let
us consider the following total Lagrangian density
\begin{equation}\label{1}
{\cal L}={\cal L}_{EM}+{\cal L}_{mat}+{\cal L}_{int}
\end{equation}
where ${\cal L}_{EM}$ is electromagnetic field Lagrangian density
\begin{equation}\label{2}
{\cal L}_{EM}=\frac{1}{2}\varepsilon_0{\bf E}^2-\frac{{\bf
B}^2}{2\mu_0}.
\end{equation}
The physical fields can be written in terms of the potentials as
${\rm E} = - \frac{{\partial {\rm A}}}{{\partial t}} - \nabla
\phi$ and ${\rm B} = \nabla \times {\rm A}$, and for simplicity we
work in the axial gauge where $\phi=0$. The dielectric medium is
modeled by a vector field ${\bf X}(\omega)$, which is suppose to
describe its electrical properties
\begin{equation}\label{3}
{\cal L}_{mat}=\int_0^\infty d\omega\frac{1}{2}{\dot{\bf
X}^2}(\omega,x)-\frac{1}{2}\omega^2{{\bf X}^2}(\omega,x).
\end{equation}
Now we define i'th component of the polarization field of the
dielectric medium as
\begin{eqnarray}\label{4}
{ P}_i(x)&=&\int_0^\infty d\omega\nu ^{(1)}(\omega){
X}_i(\omega,x)+\int_0^\infty d\omega\int_0^\infty d\omega' \nu
^{(2)}(\omega,\omega'){X}_i(\omega,x){ X}_i(\omega',x)\nonumber\\
&+&\int_0^\infty d\omega\int_0^\infty d\omega' \int_0^\infty
d\omega''\nu ^{(3)}(\omega,\omega',\omega''){ X}_i(\omega,x){
X}_i(\omega',x){ X}_i(\omega'',x)+. . .
\end{eqnarray}
In these expression, the index $i$ can take on the values $x$, $y$
and $z$ and $\nu^{(1)}, \nu^{(2)}, \nu^{(3)}, \cdots$, are the
coupling functions between the medium and the electromagnetic
field. As it can be seen from (\ref{4}), the coupling-tensor
$\nu^{(1)}$ describes the linear contribution of the interaction
and the sequence $ \nu^{(2)}, \nu^{(3)},....$ describe,
respectively, the first-order, second-order and higher orders of
non-linear interactions. The coupling-tensors $\nu^{(1)} ,
\nu^{(2)}, \nu^{(3)},...$ in (\ref{4}) are the key parameters in
this quantization scheme and we will see that the susceptibility
functions of the medium can be expressed in terms of these
coupling-tensors.

The interaction part of Lagrangian is defined by \cite{24}
\begin{equation}\label{5}
{\cal L}_{int}={\bf A}\cdot \dot{\bf P}.
\end{equation}
Now we quantize the theory using the path-integral techniques. For
this purpose let us start with the following generating functional
\cite{33}
\begin{equation}\label{6}
Z[J] = \int {\cal {D}} [\varphi] \exp{\imath\int d^4 x[{\cal
L}(\varphi (x)) + J(x)\varphi (x)]},
\end{equation}
where $\varphi$ is a scalar field and $J$ acts as a source or an
auxiliary field and different correlation functions can be found
by taking repeated functional derivatives with respect to the
field $J(x)$. The above partition function is Gaussian since the
integrand is quadratic in fields. To obtain the generating
function for the interacting fields, we first calculate the
generating function for the free fields
\begin{eqnarray}\label{7}
Z_0 [\textbf{J}_{EM},\{\textbf{J}_\omega\}] &=& \int
{\cal{D}}[{\bf A}]{\cal{D}}[{\bf X}]
\nonumber\\
&\times&\exp\big[ \imath\int d^4 x\{ {\cal{L}}_{EM}  +
{\cal{L}}_{mat} + \textbf{J}_{EM}\cdot {\bf A} +\int d\omega
\,\textbf{J}_\omega\cdot {\bf X}\}\big].
\nonumber\\
\end{eqnarray}
Using the $4$-dimensional version of Gauss's theorem
\begin{equation}\label{8}
\int d^4 x\,{\cal L}_{em} = -\int d^4 x\,\big({\bf
A}\cdot(\bigtriangledown\times\bigtriangledown\times{\bf A})+{\bf
A}\cdot\partial ^2_t{\bf A}\big),
\end{equation}
and using the integration by parts
\begin{equation}\label{9}
\int d^4 x\,\,\dot{\bf X}(\omega,x)\,\cdot\,\dot{\bf X}(\omega,x)
= -\int d^4 x\,{\bf X}(\omega,x)\,\cdot\,\frac{{\partial ^2
}}{{\partial t^2 }}\,{\bf X}(\omega,x).
\end{equation}
From Eqs. (\ref{8}) and (\ref{9}), the free generating functional (\ref{7}) can be written as
\begin{eqnarray}\label{10}
Z_0 [\textbf{J}_{EM} ,\textbf{J}_\omega ] &=& \int {\cal{D}}[{\bf
A}] {\cal{D}}[{\bf X}] \exp -\frac{\imath}{2}\big [\int d^4 x \{
({\bf A}\cdot\bigtriangledown\times\bigtriangledown\times{\bf
A}+{\bf A}\cdot\partial ^2_t{\bf A})-\textbf{J}_{EM}\cdot{\bf A}\nonumber\\
&+& \int d\omega {\bf X}(\omega,x)(\frac{{\partial ^2 }}{{\partial
t^2 }}+ \omega^2){\bf X}(x,\omega)-\textbf{J}_\omega({ x})\cdot
{\bf X}(x,\omega)\}\big]\nonumber\\
\end{eqnarray}
The integral in Eq. (\ref{10}) can be easily calculated from the
field version of the quadratic integral formula and the result is
\begin{eqnarray}\label{11}
Z_0 [\textbf{J}_{EM} ,\textbf{J}(\omega) ] & =& \exp
-\frac{i}{2}\big [\int d^4x \int d^4 x'\textbf{J}_{EM}
\cdot\G_{EM}^{(0)} (x - x')\cdot
\textbf{J}_{EM}\nonumber\\
& +&\int d^3{\bf x} \int dt  \int dt'  \int
d\omega\textbf{J}_\omega({\bf x},t)\cdot\G_{\omega}  (t -
t')\cdot\textbf{J}_\omega({\bf x},t')\big
],\nonumber\\
\end{eqnarray}
here the space component of the point $x\in \mathbb{R} ^{4}$ is
indicated in bold by ${\bf {x}}\in \mathbb{R}^{3}$ and the time
component by $t$ or $x_0\in \mathbb{R}$. The Green functions
$\G_{\omega} (t - t')$ and $\G_{EM}^{(0)} (x - x')$ are the
propagators for free fields and satisfy the following equations
\begin{equation}\label{12}
(\bigtriangledown\times\bigtriangledown\times+\,\, \frac{{\partial
^2 }}{{\partial t^2 }}\,\,) \G_{EM}^{(0)}(x - x') = \delta(x - x')
\end{equation}
\begin{equation}\label{13}
\{  \frac{{\partial ^2 }}{{\partial t^2 }} +  \omega ^2 \}
\G_{\omega\alpha,\beta}  (t - t') = \delta (t -
t')\delta_{\alpha,\beta}.
\end{equation}
Taking the Fourier transform of Eqs.(\ref{12}) and (\ref{13}) we
find
\begin{equation}\label{14}
\G^{(0)\|}_{EM,\alpha\beta}  ({\bf k},\omega) = -\frac{k_\alpha
k_\beta}{\omega ^2}
\end{equation}
and
\begin{equation}\label{14}
\G^{(0)\bot}_{EM,\alpha\beta}  ({\bf k},\omega) =
\frac{\delta_{\alpha\beta}-\widetilde{k}_\alpha
\widetilde{k}_\beta}{{\bf k}^2-\omega ^2}
\end{equation}
\begin{equation}\label{a30}
G_{\omega\alpha\beta}(\omega')=\frac{1}{\omega^2-\omega'^2-\imath
0^+}\,\delta_{\alpha\beta}
\end{equation}
where $\widetilde{k}$ is a unit vector along $\bf k$ and $\G^{\|}$
and $\G^{\bot}$ refer to the longitudinal and transverse parts of
Green's tensor in Fourier space. For further simplicity we define
\begin{eqnarray}\label{15}
{\bf J}_P=\int d\omega \nu^{(1)}(\omega) {\bf J}_\omega+\int
d\omega\int d\omega' \nu^{(2)}(\omega,\omega'){\bf J}_\omega{\bf
J}_{\omega'} +\int d\omega\int d\omega'
\nu^{(3)}(\omega,\omega',\omega''){\bf J}_\omega{\bf
J}_{\omega'}{\bf J}_{\omega''}+\cdots\nonumber\\
\end{eqnarray}
Now the generating functional of the interacting fields can be
written in terms of the free generating functional as \cite{34}
\begin{eqnarray}\label{17}
Z[{\bf J}_{EM},{\bf J}_P]& =& Z^{ - 1} [0]e^{i\int d^4
z{\cal{L}}_{int} \left(\frac{\delta } {\delta{\bf J}_{EM}
(z)},\frac{\delta }{\delta
{\bf J}_P(z)  }\right)} Z_0 [{\bf J}_{EM} ,{\bf J}_\omega ] \nonumber\\
&=&Z^{ - 1} [0] \sum\limits_{n = 0}^\infty \frac{1}{n!}\bigg
\{\imath\int d^4z\,\int d\omega_1 \,
\nu^{(1)}(\omega_1)\frac{\delta } {{\delta {\bf J}_{EM}(z) }}
\cdot \frac{\partial }{{\partial z_0}}\frac{\delta }{{\delta {\bf
J}_{\omega_1}(z) }}\nonumber\\
&+&\int d\omega_1\int d\omega_2
\,\nu^{(2)}(\omega_1,\omega_2)\frac{\delta } {{\delta {\bf
J}_{EM}(z) }} \cdot \frac{\partial }{{\partial z_0}}\frac{\delta
}{{\delta {\bf J}_{\omega_1}(z)  }}\frac{\delta }{{\delta {\bf
J}_{\omega_2}(z)  }}\nonumber\\
&+&\int d\omega_1\int d\omega_2 \int d\omega_3
\,\nu^{(3)}(\omega_1,\omega_2,\omega_3)\frac{\delta } {{\delta {\bf
J}_{EM}(z) }} \cdot \frac{\partial }{{\partial z_0}}\frac{\delta
}{{\delta {\bf J}_{\omega_1}(z)  }}\frac{\delta }{{\delta {\bf
J}_{\omega_2}(z,) }}\frac{\delta }{{\delta {\bf
J}_{\omega_3}(z)  }}\nonumber\\
&+&\cdots\,\bigg\} ^n Z_0 [{\bf J}_{EM} ,{\bf J}_\omega]
\end{eqnarray}
where $Z[0]$ a normalization factor.
\section{Linear Green tensors }
\noindent In this section we consider a linear medium, so the
sequence of nonlinear coupling functions $\nu^{(2)}$, $\nu^{(3)}$,
etc. are zero. Following the procedure presented in \cite{29}, the
linear Green's function of the electromagnetic field can be
obtained as
\begin{equation}\label{18}
\G^{(1)}_{\alpha\beta}(x - y) = i\frac{{\delta ^2 Z[J]}}{{\delta
J_{\alpha} (x)\delta J_{\beta} (y)}}\big |_{J = 0}.
\end{equation}
To facilitate the calculations we take the time-Fourier-transform
of the Green's function (\ref{18}), after some lengthy but
straightforward calculations we find the Green's function in
frequency-space as
\begin{eqnarray}\label{19}
\G^{(1)}_{EM}({\bf x} - {\bf x}',\omega)&=&\G_0 ({\bf x} - {\bf x}',\omega)\nonumber\\
&+& \int d{\bf x}_1 [\G_0 ({\bf x} - {\bf x}_1,\omega ) \int
d\omega'\{\nu ^2 (\omega' )\omega^2\G_{\omega'} (\omega
)\}\G^{(1)}_{EM} ({\bf x}_1  - {\bf x}',\omega)].
\end{eqnarray}
It can be shown that this Green's function satisfies the following
equation
\begin{equation}\label{20}
\nabla  \times \nabla  \times \G^{(1)}_{EM} ({\bf x}-{\bf
x}',\omega) - \omega ^2 \epsilon(\omega)\G^{(1)}_{EM} ({\bf x}-{\bf
x}',\omega) =
\mathord{\buildrel{\lower3pt\hbox{$\scriptscriptstyle\leftrightarrow$}}
\over \delta } ({\bf x}-{\bf x}')
\end{equation}
with the following formal solution
\begin{equation}\label{14}
\G^{(1)\|}_{EM,\alpha\beta}  ({\bf k},\omega) = -\frac{k_\alpha
k_\beta}{\omega ^2\epsilon(\omega)}
\end{equation}
\begin{equation}\label{14}
\G^{(1)\bot}_{EM,\alpha\beta}  ({\bf k},\omega) =
\frac{\delta_{\alpha\beta}-\widetilde{k}_\alpha
\widetilde{k}_\beta}{{\bf k}^2-\omega ^2\epsilon(\omega)}
\end{equation}
where $\epsilon(\omega)=1+\chi^{(1)}(\omega)$ is the linear
permittivity of the medium in frequency domain and the linear
susceptibility $\chi^{(1)}(\omega)$ can be written as
\begin{eqnarray}\label{a26/1}
\chi^{(1)}(\omega) &=& \int_0^\infty {d\omega' }\frac{[\nu^{(1)}
(\omega' )]^2}{\omega'^2-\omega^2-\imath0^+}.
\end{eqnarray}
These are complex functions of frequency which satisfy
Kramers-Kronig relations and have the properties of response
functions i.e, $\epsilon(-w^*)=\epsilon^*(\omega)$ and $Im
\epsilon(\omega)>0$ provided that
$(\nu^{(1)})^2(-\omega^*)=(\nu^{(1)})^2(\omega)$. It can be shown
that these functions have no poles in the upper half plane and
tend to unity as $\omega\longrightarrow\infty$. As a consequence,
the electric susceptibility of the medium in the time domain can
be written as
\begin{eqnarray} \label{a19}
\chi^{(1)}(t) =\left\{ \begin{array}{l}\displaystyle \int_0^\infty
{d\omega } \frac{{\sin \omega
(t)}}{\omega } [\nu^{(1)} (\omega )]^2\hspace{2cm}t > 0 \\
\\
0\hspace{+6cm}t < 0 \\
\end{array} \right.
\end{eqnarray}
which is real and causal. This proves that the constitutive
relations of the medium, with arbitrary dispersion and absorption,
is properly described by the present model. If we are given a
definite permittivity for the medium, then we can inverse the
relations (\ref{a26/1}) and find the corresponding coupling
function $\nu^{(1)}(\omega)$ as
\begin{eqnarray}\label{a27}
\nu^{(1)}(\omega) = \sqrt{\frac{{2\omega
}}{\pi}Im\chi^{(1)}(\omega)}.
\end{eqnarray}
In a similar way we can obtain the other correlation functions
among different fields \cite{29}, but here we consider only the
following correlation function
\begin{equation}\label{23}
\G^{(1)}_{EM,P}({\bf x}-{\bf
x}',\omega)=i\omega\chi^{(1)}(\omega)\G^{(1)}_{EM}({\bf x}-{\bf
x}',\omega),
\end{equation}
which shows that $\chi^{(1)}(\omega)$ is a linear electric
susceptibility \cite{30,33}. In this way, we can easily define the
nonlinear susceptibilities of the medium via two, three and
$n$-point correlation functions between electromagnetic field and
the polarization field \cite{36}. These correlation functions
satisfy the fluctuation$-$dissipation theorem \cite{28,36,37}. In
the next sections we apply the path-integral formalism to
calculate the Green's tensors and Casimir energy in the presence
of a nonlinear dielectric medium. To simplify the problem we work
in Fourier space and separate the Green's tensor to transverse and
longitudinal parts. The longitudinal part of the Green's function
does not lead to any force since this part of the Green's function
is local and the local fields do not lead to any Casimir force
\cite{29}, so we only consider the transverse part and for
simplicity drop the superscript $^\perp$.

\section{Nonlinear Green tensors }
\noindent Now let us consider the contributions induced from a
nonlinear medium. To elucidate the method, we start with the term
$\nu^{(2)}$ and generalize it to include the higher order terms
like $\nu^{(3)}$, $\nu^{(4)}$, etc. In analogue to the linear
medium we define the nonlinear Green's function of the
electromagnetic field up to the first order of nonlinearity as
\begin{equation}\label{18/b1}
\G^{(2)}_{EM}(x - y) = i\frac{{\delta ^2 Z[J]}}{{\delta J_{EM}
(x)\delta J_{EM} (y)}}\big |_{J = 0}.
\end{equation}
The nonlinear Green's function (\ref{18/b1}), after some lengthy
but straightforward calculations, can be obtained in reciprocal
space as follows
\begin{eqnarray}\label{19}
\G^{(2)}_{EM}({\bf k},\omega)&=&\G^{(1)}_{EM}({\bf k}
,\omega)+\G^{(1)}_{EM} ({\bf k},\omega
)\Delta_{NL}^{(2)}(\omega)\G^{(1)}_{EM} ({\bf k},\omega)\nonumber\\
&+& \G^{(1)}_{EM} ({\bf k},\omega
)\Delta_{NL}^{(2)}(\omega)\G^{(1)}_{EM} ({\bf
k},\omega)\Delta_{NL}^{(2)}(\omega)\G^{(1)}_{EM} ({\bf k},\omega)+ .
. .\nonumber\\
&=&\frac{1}{{\bf
k}^2-\omega^2[\epsilon(\omega)+\Delta_{NL}^{(2)}(\omega)]},
\end{eqnarray}
where
\begin{equation}\label{19/b2}
\Delta_{NL}^{(2)}(\omega)=\frac{-\imath}{2\pi}\int_0^\infty
d\omega_1 \int_0^\infty d\omega_2 \int_0^\infty
d\omega'[\nu^{(2)}(\omega_1,\omega_2)]^2
\frac{1}{\omega_1^2-(\omega-\omega')^2-\imath0^+}\times\frac{1}{\omega_2^2-\omega'^2-\imath0^+},
\end{equation}
is the first order correction to the Green's function of the
electromagnetic field due to the nonlinearity of the medium. In a
similar way, we find for $\nu^{(3)}$
\begin{eqnarray}\label{20}
\G^{(3)}_{EM}({\bf k},\omega)&=&\frac{1}{{\bf
k}^2-\omega^2[\epsilon(\omega)+\Delta_{NL}^{(2)}(\omega)+\Delta_{NL}^{(3)}(\omega)]}
\end{eqnarray}
where
\begin{eqnarray}\label{21}
\Delta_{NL}^{(3)}(\omega)&=&\big(\frac{-\imath}{2\pi}\big)^2\int_0^\infty
d\omega_1 \int_0^\infty d\omega_2 \int_0^\infty d\omega_3
\int_0^\infty d\omega_2' \int_0^\infty
d\omega_3'[\nu^{(3)}(\omega_1,\omega_2,\omega_3)]^2\nonumber\\
&&\times\frac{1}{\omega_1^2-(\omega-\omega'_2)^2-\imath0^+}
\times\frac{1}{\omega_2^2-\omega'^2_2-\imath0^+}
\times\frac{1}{\omega_2^2-\omega'^2_3-\imath0^+}
\end{eqnarray}
following these procedure we find the nonlinear Green's function
of the electromagnetic field up to the n'th order of the
nonlinearity as
\begin{eqnarray}\label{22}
\G^{(n)}_{EM}({\bf k},\omega)&=&\frac{1}{{\bf
k}^2-\omega^2[\epsilon(\omega)+\Delta_{NL}^{(2)}(\omega)+\Delta_{NL}^{(3)}(\omega)+\cdots+\Delta_{NL}^{(n)}(\omega)]}
\end{eqnarray}
where
\begin{eqnarray}\label{23}
\Delta_{NL}^{(n)}(\omega)&=&\big(\frac{-\imath}{2\pi}\big)^{n-1}\int_0^\infty
d\omega_1 \int_0^\infty d\omega'_2\int_0^\infty d\omega_2
 \cdots \int_0^\infty
d\omega'_{n}\int_0^\infty d\omega_{n} [\nu^{(n)}(\omega_1,\omega_2,\cdots,\omega_{n})]^2\nonumber\\
&&\times\frac{1}{\omega_1^2-(\omega-\omega_2'-\cdots-\omega_{n}')^2-\imath0^+}
\times\frac{1}{\omega_2^2-\omega_2'^2-\imath0^+}\times\cdots\times
\,\frac{1}{\omega_{n}^2-\omega_{n}'^2-\imath0^+}.\nonumber\\
\end{eqnarray}
In a similar way we can obtain correlation functions among
different fields. For example, 3-points Green's function \cite{35}
up to the first order of nonlinearity is defined by
\begin{eqnarray}\label{24}
\G^{(2)}_{EM,EM,P}({\bf x}-{\bf x}',{\bf x}_1-{\bf
x}',\omega_1,\omega_2)&=&\frac{{\delta ^2 Z[J]}}{{\delta J_{EM}
(x)\delta J_{EM} (x_1)\delta J_{\bf P} (x')}}\big |_{J = 0}.\nonumber\\
&=&- \omega_1 \omega_2 \chi^{(2)}(\omega_1,\omega_2)
\G^{(1)}_{EM}({\bf x}-{\bf
x}',\omega_1)\G^{(1)}_{EM}({\bf x}_1-{\bf x}',\omega_2)\nonumber\\
\end{eqnarray}
where
\begin{equation}\label{25}
\chi^{(2)}(\omega_1,\omega_2)=\int_0^\infty d\omega'_1
\int_0^\infty d\omega'_2
\nu^{(2)}(\omega'_1,\omega'_2)\nu^{(1)}(\omega'_1)\nu^{(1)}(\omega'_2)
\frac{1}{{\omega'}_1^2-{\omega}^2_1-\imath0^+}
\times\frac{1}{{\omega'}_2^2-{\omega}^2_2-\imath0^+},
\end{equation}
is the second order susceptibility. Also we note that in deriving
Eq.(\ref{24}) only leading terms have been kept and very small
terms have been ignored. In order to make contact with standard
notation, let us recall the definition of the second order of the
polarization within the framework of response theory
\begin{equation}\label{26}
{ { \bf P}^{(2)}}(t) = \int_{-\infty}^{t}  \int_{-\infty}^{t}
{dt\,}dt' \, \chi^{(2)} (t - t',t-t'')  {\bf E}(t'){\bf E}(t')\, +
{ \bf P}_{N}^{(2)} (t),
\end{equation}
where ${ \bf P}_{N}^{(2)}$, is the noise operator up to the second
order of nonlinearity. It is easy to show that Eq.(\ref{25}) in
time-domain can be written as
\begin{equation}\label{26/1}
\chi^{(2)}(t_1,t_2) = \int_0^\infty  {d\omega_1 } \int_0^\infty
{d\omega_2 }\nu^{(2)} (\omega_1,\omega_2 )
\nu^{(1)}(\omega_1)\nu^{(1)}(\omega_2)\frac{{\sin \omega_1
t_1}}{\omega _1}\times\frac{{\sin \omega_2 t_2}}{\omega_2 },
\end{equation}
which is consistent with the results have been reported in
\cite{24}. Following these procedure we find the susceptibility up
to the $n$'th order as follows
\begin{eqnarray}\label{27}
\chi^{(n)}({\omega}_1,\cdots,{\omega}_n)&=&\int_0^\infty d\omega'_1
\cdots \int_0^\infty d\omega'_n
\nu^{(n)}(\omega'_1,\cdots,\omega'_n)\nu^{(1)}(\omega'_1)\cdots\nu^{(1)}(\omega'_n)\nonumber\\
&&\times\frac{1}{{\omega'}_1^2-{\omega}^2_1-\imath0^+} \times
\cdots \times \frac{1}{{\omega'}_n^2-{\omega}^2_n-\imath0^+},
\end{eqnarray}
which in time-domain can be expressed as
\begin{eqnarray}\label{27/1}
\chi^{(n)}(t_1,\cdots,t_n)&=&\int_0^\infty d\omega_1 \cdots
\int_0^\infty d\omega_n
\nu^{(n)}(\omega_1,\cdots,\omega_n)\nu^{(1)}(\omega_1)\cdots\nu^{(1)}(\omega_n)\nonumber\\
&&\times\frac{{\sin \omega_1 t_1}}{\omega _1}\times\cdots
\times\frac{{\sin \omega_n t_n}}{\omega _n}.
\end{eqnarray}
Now if we are given definite $n$th order susceptibility of the
medium then we can inverse the relations (\ref{27/1}) and using
Eq.(\ref{a27}) the corresponding coupling function $\nu^{(n)}$ can
be found as
\begin{equation}\label{27/2}
\nu^{(n)}(\omega_1,\omega_2,\cdots,\omega_n)=\frac{
\sqrt{\omega_1\omega_2\cdots\omega_n}\,\, Im [
\chi^{(n)}({\omega}_1,{\omega}_2,\cdots,{\omega}_n)]}{(2\pi)^{n/2}\sqrt{Im[
\chi^{(1)}(\omega_1)]  Im[ \chi^{(1)}(\omega_2)] \cdots Im[
\chi^{(1)}(\omega_n)]}},
\end{equation}
where
\begin{eqnarray}\label{27/3}
Im \chi^{(n)}({\omega}_1,\cdots,{\omega}_n)&=&\int_0^\infty dt_1
 \cdots \int_0^\infty dt_n \,\chi^{(n)}(t_1,\cdots,t_n)\,{\sin
\omega_1 t_1}\cdots \,{\sin \omega_n t_n}\nonumber\\
&=&\frac{\omega_1\omega_2\cdots\omega_n}{(2\pi)^n}\int_0^\infty
d\omega'_1 \cdots \int_0^\infty d\omega'_n
\chi^{(n)}({\omega'}_1,\cdots,{\omega'}_n)\nonumber\\
&&\times\frac{1}{\omega_1^2-{\omega'}^2_1+\imath0^+} \times \cdots
\times \frac{1}{\omega_n^2-{\omega'}^2_n+\imath0^+}.
\end{eqnarray}
From Eq.(\ref{27/2}) it is clear that the $n$th order coupling
function is an odd function in frequencies. Therefore, the $n$th
order coupling function will be an even(odd) function if $n$ is
even(odd). Using this symmetry, Eq.(\ref{23}) can be rewritten as
\begin{eqnarray}\label{27/4}
\Delta_{NL}^{(n)}(\omega)&=&\frac{2\pi\imath}{(8\pi)^{n}}\int_0^\infty
d\omega_2  \cdots \int_0^\infty d\omega_{n}  \frac{
 \big(Im [\chi^{(n)}(\omega-\omega_2-\cdots-\omega_{n},{\omega}_2,\cdots,{\omega}_n)]\big)^2}{{Im[
\chi^{(1)}(\omega-\omega_2-\cdots-\omega_{n})]Im[
\chi^{(1)}(\omega_2)] \cdots Im[ \chi^{(1)}(\omega_n)]}}.\nonumber\\
\end{eqnarray}

\section{Calculating the Casimir force}
\subsection{General formalism}
\noindent In this section we calculate the Casimir force for the
simplest configuration consisting of two prefect conducting plates
separated by a nonlinear medium of width $h$. For this purpose we
consider the electromagnetic field as transverse magnetic(TM) and
transverse electric(TE) modes. These modes are represented by
scaler fields which satisfy the Dirichlet
\begin{equation}\label{28}
\varphi (X_\alpha) = 0
\end{equation}
or Neumann
\begin{equation}\label{29}
{\partial_n}\varphi (X_\alpha) = 0
\end{equation}
boundary conditions where $X_\alpha, (\alpha=1,2)$ is an arbitrary
point on the conducting plates. To obtain the partition function
from the Lagrangian we apply the Wick's rotation, $(t\rightarrow
\imath\tau)$ and change the signature of space-time from Minkowski
to Euclidean. The Dirichlet or Neumann boundary conditions can be
represented by auxiliary fields $\psi _\alpha (X_\alpha )$ as
\cite{38}
\begin{equation}\label{30}
\delta (\varphi (X_\alpha  )) = \int {\cal {D}}[\psi_\alpha
(X_\alpha )]e^{\imath\int \psi (X_\alpha  )\varphi (X_\alpha  )},
\end{equation}
and
\begin{equation}\label{31}
\delta ({\partial_n}\varphi (X_\alpha  )) = \int {\cal
{D}}[\psi_\alpha (X_\alpha )]e^{\imath\int {\partial_n}\psi
(X_\alpha )\varphi (X_\alpha  )}.
\end{equation}
Using Eqs.(\ref{30},\ref{31}) the partition function can be
written as
\begin{equation}\label{32}
 Z_D = Z_0^{ - 1} \int {\cal {D}}[\varphi] \prod\limits_{^{a = 1} }^2
  {\cal D}[\psi_\alpha (X_\alpha  )])e^{S_D[\varphi ]}
\end{equation}
\begin{equation}\label{33}
 Z_N = Z_0^{ - 1} \int {\cal {D}}[\varphi] \prod\limits_{^{a = 1} }^2
  {\cal D}[\psi_\alpha (X_\alpha  )])e^{S_N[\varphi ]}
\end{equation}
where $S_D (\varphi )$ and $ S_N (\varphi )$ are defined
respectively  by
\begin{equation}\label{34}
S_D [\varphi ] = \int d^{(n+1)} x\{{\cal L} (\varphi(x))+ \varphi (x)\sum\limits_{\alpha = 1}^2 \int d^{(n)}
X\delta (X - X_\alpha )\psi _\alpha  (x)\}.
\end{equation}
and
\begin{equation}\label{35}
S_N [\varphi ] = \int d^{(n+1)} x\{{\cal L} (\varphi(x))+ \varphi
(x)\sum\limits_{\alpha = 1}^2 \int d^{(n)} X\delta (X - X_\alpha
)\partial_n\psi _\alpha  (x)\}.
\end{equation}
By comparing the Eqs.(\ref{34}, \ref{35}) and (\ref{6}), we can
rewrite Eqs. (\ref{34}, \ref{35}) as
\begin{equation}\label{36}
Z_D  = \int \prod_{\alpha=1}^{2}{\cal {D}}[\psi_\alpha(x)] Z(\sum\limits_{\alpha  = 1}^2  \int {d^nX \delta (X -
X_\alpha )} \psi _\alpha (X))
\end{equation}
and
\begin{equation}\label{37}
Z_N  = \int \prod_{\alpha=1}^{2}{\cal {D}}[\psi_\alpha(x)]
Z(\sum\limits_{\alpha  = 1}^2  \int {d^nX \delta (X - X_\alpha )}
\partial_n\psi _\alpha (X))
\end{equation}
where $Z(\sum\limits_{\alpha  = 1}^2  \int {d^nX_\alpha \delta (X
- X_\alpha  )} \psi _\alpha (X))$ and $Z(\sum\limits_{\alpha  =
1}^2  \int {d^nX_\alpha \delta (X - X_\alpha  )}\partial_n\psi
_\alpha (X))$ are the generating functionals of interacting fields
defined in Eq.(\ref{17}) with imaginary time. From Eqs.(\ref{36},
\ref{37}) and (\ref{18}) the respective partition functions can be
written as
\begin{equation}\label{38}
Z_D  = \int \prod\limits_{\alpha  = 1}^2  {\cal {D}}[\psi _\alpha
(X_\alpha )]e^{\imath S_D (\psi _1 ,\psi _2 )},
\end{equation}
and
\begin{equation}\label{39}
Z_N  = \int \prod\limits_{\alpha  = 1}^2  {\cal {D}}[\psi _\alpha
(X_\alpha )]e^{\imath S_N (\psi _1 ,\psi _2 )},
\end{equation}
where
\begin{equation}\label{40}
\imath S_D (\psi _1 ,\psi _2 ) = \imath J_D\, {\cal G}\,J_D,
\end{equation}
and
\begin{equation}\label{41}
\imath S_N(\psi _1 ,\psi _2 ) = \imath J_N\, {\cal G}\,J_N,
\end{equation}
and $J_D(X)$ and $J_N(X)$ are respectively defined by
\begin{equation}\label{42}
J_D(X)=\int {d^{n}X\, \delta (X - X_\alpha  )}\, \psi _\alpha (X),
\end{equation}
\begin{equation}\label{43}
J_N(X)=\int {d^{n}X\, \delta (X - X_\alpha  )}\, \partial_n\psi
_\alpha (X).
\end{equation}
We will define the function ${\cal G}$ shortly. The partition
functions defined by (\ref{38}) and (\ref{39}) are calculated
straightforwardly  \cite{38}
\begin{equation}\label{44}
Z_D  = \frac{1}{{\sqrt {\det \Gamma_D (x,y,h)} }},
\end{equation}
and
\begin{equation}\label{45}
Z_N  = \frac{1}{{\sqrt {\det \Gamma_N (x,y,h)} }},
\end{equation}
where
\begin{equation}\label{46}
\Gamma_D (x,y,h) =\big [\begin{array}{*{20}c}
   {{\cal G}(x - y,0)} & {{\cal G}(x - y,h)}  \\
   {{\cal G}(x - y,h)} & {{\cal G}(x - y,0)}  \\
\end{array}\big ]
\end{equation}
and
\begin{equation}\label{47}
\Gamma_N (x,y,h) =\big [\begin{array}{*{20}c}
   -\partial^2_z{{\cal G}(x - y,0)} & -\partial^2_z{{\cal G}(x - y,h)}  \\
   -\partial^2_z{{\cal G}(x - y,h)} & -\partial^2_z{{\cal G}(x - y,0)}  \\
\end{array}\big ]
\end{equation}
where ${\cal G}$ is the Green's function of the fields after a
Wick's rotation. Now in order to calculate the Casimir force we
define the effective action as
\begin{equation}\label{48}
S_{eff}  =- \imath \ln Z_D [h]
\end{equation}
from which the Casimir force can be obtained easily as
\begin{equation}\label{49}
F = \frac{{\partial S_{eff} (h)}}{{\partial h}}.
\end{equation}
It is easy to show that the Dirichelet and Neumann boundary
condition are formally the same and when the plates are complete
conductors lead to the same result. To save brevity in what
follows we only calculate Casimir force for Dirichelet boundary
conditions and its generalization to Neumann boundary conditions
is straightforward.

\subsection{Linear and nonlinear effects of the medium (T=0)}
\noindent To calculate the Casimir energy in presence of a
nonlinear medium boundary conditions are imposed on the
electromagnetic field and the $\Gamma$ tensor can be written as
\begin{equation}\label{50}
\Gamma_{EM,EM} (x,y,h) =\big [\begin{array}{*{20}c}
   {{\cal G}_{EM,EM}(x - y,0)} & {{\cal G}_{EM,EM}(x - y,h)}  \\
   {{\cal G}_{EM,EM}(x - y,h)} & {{\cal G}_{EM,EM}(x - y,0)}  \\
\end{array}\big ].
\end{equation}
Now to obtain the Casimir force we proceed in Fourier-space since
the $\Gamma$ tensor is diagonal in this space. The Fourier
transformation of ${{\cal G}_{EM,EM}(x - y,h)}$ is
\begin{eqnarray}\label{51}
{{\cal G}_{EM,EM}(p,q,h)}&=&\int dx\int dy e^{\imath p\cdot x+\imath
q\cdot y}G_{EM,EM}(x-y,h)\nonumber\\
&=&\frac{e^{-h\sqrt{[ \epsilon(\imath p_0 ) +
\Delta_{NL}^{(2)}(\imath p_0)+\cdots+\Delta_{NL}^{(n)}(\imath
p_0)]p_0^2+p_1^2+p_2^2}}} {2\sqrt{[ \epsilon(\imath p_0 ) +
\Delta_{NL}^{(2)}(\imath p_0)+\cdots+\Delta_{NL}^{(n)}(\imath
p_0)]p_0^2+p_1^2+p_2^2}}(2\pi)^3\delta(p+q),\nonumber\\
\end{eqnarray}
where $p = (p_0,{\bf p})$ and $\bf p$ is a vector parallel to the
conductors, $p_0$ is the temporal component of $p$. Thus, in this
case, the Casimir force is
\begin{equation}\label{52}
F_c  = -\int \frac{{d^3 p}}{{(2\pi )^3 }}[\frac{{\cal
Q}(p)}{{e^{2{{\cal Q}(p)}h}  - 1}}],
\end{equation}
where
\begin{equation}\label{53}
{\cal Q}(p) = \sqrt {{\bf p}^2+p_0^2[ \epsilon(\imath p_0 ) +
\Delta_{NL}^{(2)}(\imath p_0)+\cdots+\Delta_{NL}^{(n)}(\imath
p_0)]}.
\end{equation}
It is easy to show that the Casimir force in the presence of a
nonlinear medium (\ref {52}) is similar to the Casimir force in
the presence of a linear one, the only difference is in the
definition of ${\cal Q}(p)$. In the absence of the nonlinearity
i.e.,
$\Delta_{NL}^{(2)}(\omega)=\cdots=\Delta_{NL}^{(n)}(\omega)=0$,
the original Casimir force between two plates in the presence of a
linear medium is recovered.

\subsection{Finite temperature}
\noindent Our considerations so far have been restricted to zero
temperature. In fact the temperature corrections to the Casimir
force turned out to be negligible in experiments \cite{9},
\cite{39}-\cite{40} where the measurements were performed in the
separation range $h<1 mm$. But, at $h>1 mm$, as in \cite{8}, the
temperature corrections make larger contributions to the
zero-temperature force between perfect conductors. The
generalization of the formalism to this case is straightforward.
The inclusion of temperature may be done in the usual manner
\cite{3}, \cite{41}-\cite{43}. The finite temperature expression,
can be found by replacing the frequency integral by a sum over
Matsubara frequencies according to the rule
\begin{equation}\label{a49}
\hbar \int_0^\infty  {\frac{{d\xi }}{{2\pi }}} f(\iota \xi)\,\,\,
\to \,\,\,k_B T\sum\limits_{l = 0}^{\infty\,\,\,'}  {f(\iota \xi_l
),\,\,\,\,\,\,\,\,\,\,\,\,\,\,\,\,\,\,\,\,\,\,}  \xi_l  = {{2\pi
k_B Tl} \mathord{\left/
 {\vphantom {{2\pi k_B Tl} \hbar }} \right.
 \kern-\nulldelimiterspace} \hbar }
\end{equation}
where $T$ and $k_B$ are the temperature and Boltzmann constants
and the prime over the summation means the zeroth term should be
given half weight as is conventional. Therefore the Casimir force
at finite temperature can be expressed as
\begin{equation}\label{52}
F_c  = -\frac{k_B T}{4\pi^2\hbar}\sum\limits_{l =
0}^{\infty\,\,\,'}\int {d^{\,2} {\bf p}}[\frac{{\cal
Q}(p)}{{e^{2{{\cal Q}(p)}h}  - 1}}]
\end{equation}

\section{Conclusion}
\noindent Based on a canonical approach and using path-integral
techniques, electromagnetic field in a nonlinear dielectric is
quantized and Casimir force, in the presence of a nonlinear
medium, at finite temperature, is calculated.

\end{document}